\documentclass[aps,prc,twocolumn,amsfonts,amsmath,amssymb,showpacs,floatfix]{revtex4-1}

\usepackage{graphicx}
\usepackage{hyperref}
\hypersetup{colorlinks,citecolor=blue,filecolor=blue,linkcolor=blue,urlcolor=blue}
\usepackage{color}

\begin{document}

\title{Single-Nucleon Energies Changing with Nucleon Number}

\author{J.~P. Schiffer}
\email{schiffer@anl.gov}
\affiliation{Physics Division, Argonne National Laboratory, Lemont, Illinois 60439, USA}
\author{B.~P.~Kay}
\affiliation{Physics Division, Argonne National Laboratory, Lemont, Illinois 60439, USA}
\author{J.~Chen}
\affiliation{Physics Division, Argonne National Laboratory, Lemont, Illinois 60439, USA}

\date{\today}

\begin{abstract}

The broad range of accumulated experimental data on the binding energies for single-particle states in nuclei is examined as a function of the constituent number of neutrons and protons and an unexpectedly simple pattern emerges. The dependence of the energies of neutron states on the number of constituent protons, or of proton states on the number of neutrons, are very similar to each other and the sign reflects the well-known strong attraction. For the same kind of nucleons changing as in the state -- energies for neutron states with neutron number changing or proton states with protons -- the dependence is at least a factor of four weaker in magnitude and slightly repulsive, except when the changing nucleons are only within the same orbit as the state. The systematics of the accumulated data are presented with a minimum of use made of model assumptions. 

\end{abstract}

\maketitle

The representation of the structure of nuclei in terms of the shell-model~\cite{Mayer49,Jensen49} has been enormously successful over the past $\sim$70 years, providing the framework for calculating nuclear structure. In this approximation nucleons are contained in a mean field in relatively unperturbed single-particle orbits. The mean field is taken to be provided by the rest of the nucleons, and the structure is dependent on the remaining residual effective interaction between the valence nucleons.

In this work we put together experimental information on the binding energies of single-particle (s.p.)\ states in nuclei, formed by adding or removing a valence nucleon from an even nucleus. These are often referred to as effective single-particle energies (ESPEs), as defined, for instance in Otsuka {\it et al.}~\cite{Otsuka21}. In this work, we are talking of energies for states or centroids of single-particle strength in the case of fragmentation~\cite{French66}, and to avoid repeating various different nomenclatures, we define $E_j^\pi$ and $E_j^\nu$ to stand for the experimental energies of proton and neutron ESPEs with quantum numbers $j$ (always as binding energy), and sometimes just refer to them as `states'.

We focus on how these energies change as a function of the number and kind of constituent nucleons that make up the rest of the nucleus. This information provides a relatively model-independent measure of the characteristics of both the mean field and the effective interaction that underlies the shell model.

There have been a number of related treatments (for example~\cite{Cohen63}) but the focus was mostly on obtaining the parameters of a potential describing the data. Here we survey the available data and attempt to examine systematics in as model-independent a way as seems practical.  Two papers by Bansal and French~\cite{Bansal64} are also relevant antecedents of this approach.

\begin{figure}[h!]
\centering
\includegraphics[scale=0.44]{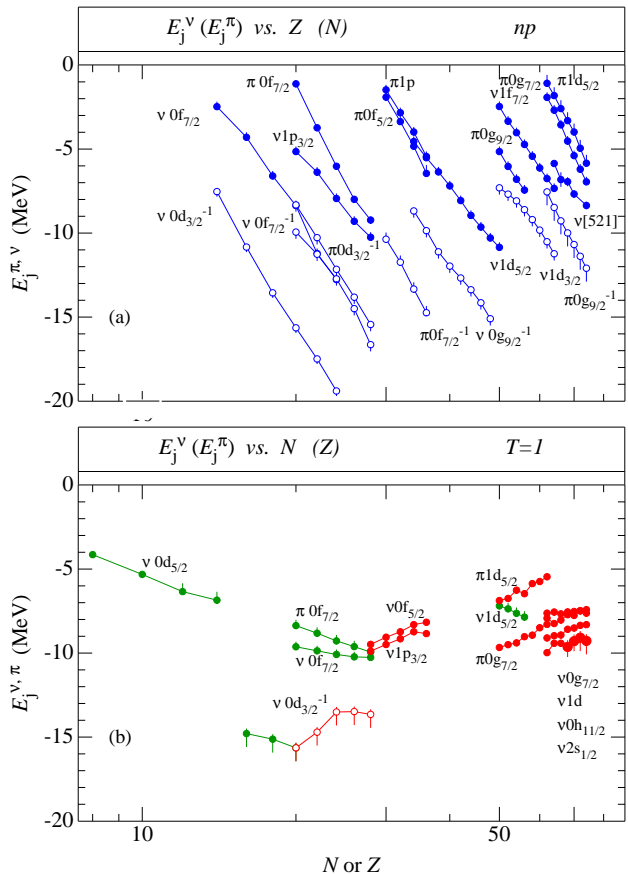}
\caption{\label{fig1} Effective energies ($E_j^{\pi,\nu}$) plotted in terms of absolute binding energies on the target as a function of the changing number of protons or neutrons. The points corresponding to one state are connected by lines. Each segment is labeled as to the configuration of the state studied. Data from purely hole states are open circles, others are filled symbols. In (a), the changing nucleons are of the other kind to the state tracked ($np$ interaction) and in (b), the changing nucleons are the same kind as in the state studied (the interaction is $T=1$) segments where $only$ the occupancy of the state in question is changing are shown in green, those where occupancies are changing in several orbitals for the segment are in red. For the data in panel (b) an approximate Coulomb correction is applied to the proton states.}
\end{figure}

The data considered are presented in Fig.~\ref{fig1}. The figure represents experimental knowledge published over the past $\sim$60 years, much of it has been analyzed in the context of the particular nuclei that were studied, and have not been compared to each other in a consistent way since the early attempts that were mentioned above. The bulk of the data that is easily interpreted was used, there is certainly more, but it will require careful consideration of the slightly different assumptions made in each experiment to extract the relevant information consistently. Some care was taken that no data sets were omitted, just because they did not fit the pattern, in fact, Fig.~\ref{fig1}(b) shows how a relatively minor feature may have lead to some physical insight, these are colored red and green, and discussed in more detail below.

The data used, energies, angular momenta, and nucleon numbers, are close to being observables, independent of models. They are published (for example, Refs.~\cite{Elbek69,Wildenthal71,Booth75,Schiffer04,Schiffer13,Szwec21}) and in compilations~\cite{nudat} with some specific details given in the Supplemental Material~\cite{supmat}.

In total, thirty-three segments of single-particle energies were used, a total of about 200 points. In some cases, the energies of the likely dominant component of a state is known, but the strength of the nucleon adding or removing reaction is not, and assumptions were made. In others, there are several measurements that do not quite agree. Most of these differences are consistent with the uncertainties shown in the estimated errors in the figures and do not significantly impact the general features discussed here. Most of the nuclei included are in spherical regions, with one exception, where careful transfer data in a region of well-deformed nuclei exists, following a particular state through some 13 nuclei with transfer reactions. 

As an example, the first segment on the left of the upper panel of Fig.~\ref{fig1}, the hollow points correspond  to the energies of the $0d_{3/2}$ neutron-hole excitations in $N=20$ nuclei, in $^{36}$S, $^{38}$Ar, $^{40}$Ca, $^{42}$Ti,  and $^{44}$Cr. There are reaction data available for all but the last two points, where assumptions were made, and the isobaric analog state has to be considered in $^{37}$Ar. Some of such details are discussed in the Supplemental Material~\cite{supmat}.

In the simple cases, where there is a single neutron or proton (or hole) on a closed shell of the same type of nucleon, and the other type of nucleon is varied, the nucleon-adding or removing strength is often concentrated in one state. Where details of  nuclear structure may be more complicated, and the s.p.\ degree of freedom is fragmented, the spectroscopic-factor weighted centroid energies are taken~\cite{French66}. Hole states are treated on the same basis as particle states. The energies are with respect to zero binding, thus always negative.

These segments in Fig.~\ref{fig1}(a) are approximately parallel and moving downward, becoming more bound, with added nucleons. The horizontal scale is logarithmic. Where measured, the segments for particle and hole states are almost contiguous.  Some of the minor variations in slope are caused by the specific $j$-dependence of the tensor force, that has been investigated extensively in connection with changing shell structure~\cite{Otsuka05}.

Fig.~\ref{fig1}(b) shows the data, where the $same$ species of nucleons are varied as the ESPEs tracked, the values for $E_j^{\pi,\nu}$ are derived by combining the energies of particle and hole excitations, taken as the energy centroids of the nucleon-adding and -removing strengths as defined by Baranger~\cite{Baranger70}: $E_j=E_j^+G_j^++E_j^-G_j^-$. Here $ E_j'^{\pm}$ are the centroid energies for adding or removing a nucleon, and $G^{\pm}$ are the summed strengths, with  $G_j^++G_j^-=1.0$.

This procedure has to be used only when the nucleons varying are of the same species as that of the state studied, and their occupancy may be changing either along with those for other orbitals, as in the strings of Ni or Sn isotopes for neutron states when $N$ is changing, or $only$ the occupancy of that orbital, as for  $\nu0f_{7/2}$ in the Ca isotopes. 

The slopes of all these segments are much less steep than those in Fig.~\ref{fig1}(a), and are shown in different colors, depending on whether only nucleons in the same orbit as the state followed are changing, or the occupancies of several are changing. They are clearly different. 

\begin{table}
\caption{\label{tab1} The logarithmic derivative of the $E_j^{\pi,\nu}$ with respect to nucleon number.}
\newcommand\T{\rule{0pt}{2.5ex}}
\newcommand \B{\rule[-1.5ex]{0pt}{0pt}}
\begin{ruledtabular}
\begin{tabular}{lr}
\T\B Slope & $dE_j^{\pi,\nu}/d({\rm ln}N)$ (MeV) \\
\hline
\T Average ({\it np}) &  $-$21.3 (3.7) \\
\T Average ($T=1$), one orbit &  $-$4.1 (1.3)\\
\T Average ($T=1$), several orbits &  $+$4.6 (1.5)\\
\end{tabular}
\end{ruledtabular}
\end{table}

Since the segments shown in Fig.~\ref{fig1}(b) involve one type of nucleon, only the $T=1$ interaction can play a role in the data. An approximate correction to remove Coulomb effects (discussed in the Supplemental Material~\cite{supmat}) has been applied to the cases where $Z$ is changing for proton states in order to allow a meaningful comparison between neutron and proton data, but higher-order Coulomb effects (e.g.\ changing radius because of added nucleons) are complex, and no corrections were made. 

In one case, the neutron $0d_{3/2}$ holes in the Ca isotopes, the sequence of energies shows a break at $N=20$, where this orbit is filled. The sequence is split at the shell closure, the data below $N=20$ are separated from the data above. While the $0d_{3/2}$ orbit is filling, the slope shows the characteristic downward trend, as in all the cases where only the orbit considered is changing its occupancy. Above $N=20$, as the $0f_{7/2}$ orbit fills, the slope is upward, as is always the case when the occupancy of more than the state studied is changing its occupancy. This is the only case of a sequence spanning two regions.

From Fig.~\ref{fig1}(a) it is clear that the influence of added neutrons on proton states, or of added protons on neutron states is strongly attractive.  This feature has been qualitatively recognized for a long time and is certainly recognized in the semi-empirical mass formula~\cite{Weizsacker35,Bethe36}, where the appearance of an asymmetry term in the expression reflects this. The effective interaction between a neutron and proton  is sometimes referred to as the $np$ interaction, which in the isospin nomenclature amounts to the average of the $T=0$ and $T=1$ interactions between two nucleons. In comparison, the energy changes much less for neutron states as neutrons are added, or for protons with $Z$ changing, where the interaction must be isospin $T=1$ by definition.  Thus, as is well known and expected from one-pion exchange, the $T=0$ interaction is much stronger than the $T=1$. To study the features in the data further we analyze the slopes of the line segments. Note that one case is included, that strictly speaking does not quite belong, but is the only well-deformed case: 
the Nilsson orbit $1/2^-[521]$ had been extensively studied experimentally in transfer reactions and the approximate dependence on proton number was plotted, while not strictly constraining the number of neutrons.

\begin{figure}
\centering
\includegraphics[scale=0.38]{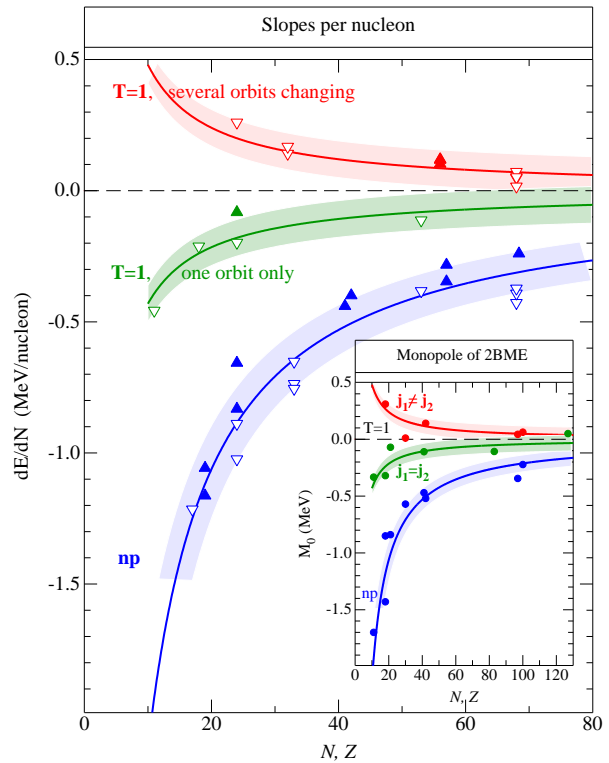}
\caption{\label{fig2} The slope of effective energies vs. nucleon number for the line segments shown in Fig.~\ref{fig1}.  Blue dots are the slopes for segments where the nucleons changing are different from the one in the state, $np$,  those in panel (a) of Fig.~\ref{fig1}. The  red and green points correspond to the slopes of segments where the nucleons are the same species as in the state, $T=1$, from panel (b)  in the previous figure. The lines represent a 1/$N$ or 1/$Z$ dependence, normalized to the data (the assumption that the slope in fractional change in the numbers of nucleons is constant).  The shaded areas indicate the approximate range of the data. The inset shows the monopole of the empirical two-body matrix elements from Ref.~\cite{Schiffer76} as discussed in the text with the nucleon numbers the average of the species contributing to that multiplet for that nucleon. The shaded areas and the lines are the same in the inset as in the main figure.}
\end{figure}

The apparent similarity between the slopes of energies changing with nucleon number, is a feature of plotting them on a logarithmic scale in $N$ or $Z$. The pattern would be quite different, were the data plotted on a linear scale. The slope for each segment, $d(E_j^{\pi,\nu})/ dN$ or $dZ$, are plotted in Fig.~\ref{fig2}. The lines correspond to constant change in energy for the same $fractional$ change in nucleon number, consistent with the naive picture that each nucleon contributes equally to the overall binding field. 

Indeed, the slopes of the different line segments follow the lines with some scatter, likely from differences in overlaps between orbits and the details of the effective interaction, such as the tensor component. But the scatter is small compared to its magnitude.

The mean values of the slopes (logarithmic derivative with respect to nucleon number) in the various subgroups and the rms variation among them are shown in  Table~\ref{tab1}. The individual values for these are given in the Supplemental Material~\cite{supmat}.

As is evident from Table~\ref{tab1}, where the changing nucleons are the same as the one tracked ($T=1$), the slope is less in absolute magnitude than for the $np$ case. This is the most dramatic difference. It is approximately consistent with the empirical `symmetry term' in the Semi-Emiprical Mass Formula and the Woods Saxon potentials of the optical model for nucleons which is briefly discussed further below. But to our knowledge, the universality of this behavior in the data for binding energies of single-particle states has not been previously noted. The difference in the two groups of $T=1$ segments is also evident in the figure, and is seen in Table~\ref{tab1}, with the scatter within each of the two subgroups (1.4 and 1.7~MeV), much less than the separation between them (9.1~MeV).

In a past survey, some of us studied two-nucleon spectra outside closed shells~\cite{Schiffer76}, where complete multiplets for two nucleons outside doubly-closed-shell nuclei were reviewed. At doubly-closed shells, such as $^{16}$O, $^{40}$Ca, $^{90}$Zr, or $^{208}$Pb, where the difference in binding energies of one particle outside the core nucleus is compared with that of the two nucleons, yielding the complete multiplet for the two-body ($NN$) interaction in nuclei such as $^{210}$Bi, $^{210}$Pb, and $^{208}$Bi.

The monopole moment $M_0$  of the experimental multiplets should be closely linked to (effectively the same as) the slopes derived here. This is shown in the inset of Fig.~\ref{fig2}. The energy in each data point in this inset is the monopole moment of a multiplet from Ref.~\cite{Schiffer76}, with the nucleon number the average of that for the two nucleon species contributing. The process of going from the binding energy of one nucleon to that of two nucleons outside the core, amounts essentially to the same as determining a slope with only the first step after a closed shell used.  

It is remarkable that the $M_0$ values for identical orbits, fall mostly below 0, and the ones for non-identical ones (red) above it. Thus the observed patterns in the slopes is confirmed, and not an artifice arising from the approximations in of the Baranger interpolation~\cite{Baranger70}, for instance. We will return to this briefly below.

The purpose of this note was primarily to present the data in as model-independent way as possible. Nevertheless, some discussion of the potential for the nucleon-nucleus interaction seems appropriate. This has been approximated in shell-model studies of nuclear structure by oscillator potentials for the sake of calculational convenience, and more recently by {\it ab initio} calculations where the interaction is derived from the empirical free $NN$ scattering. In parallel, nucleon scattering was parameterized historically since the introduction of the empirical optical model~\cite{Feshbach54}, to fit data on the scattering of nucleons on nuclei (for example, Ref.~\cite{Koning03}) and later also used to describe bound single-particle states adjacent to closed shells~\cite{Schwierz07}.  To our knowledge such potentials have not been applied systematically  to the change in bound-state energies as a function of nucleon number. The functional form has a diffuse edge and finite binding, and is characterized by radius that varies as $r_0A^{1/3}$ and a strength that is a constant, $V_0$, plus a symmetry term proportional to $\pm V_{sym}(N-Z)/A$. The exact parameters do not matter for this qualitative behavior, the values used were $r_0=1.17$ fm, $V_0=-50$ MeV, and $V_{sym}=30$ MeV.

\begin{figure}
\includegraphics[scale=0.33]{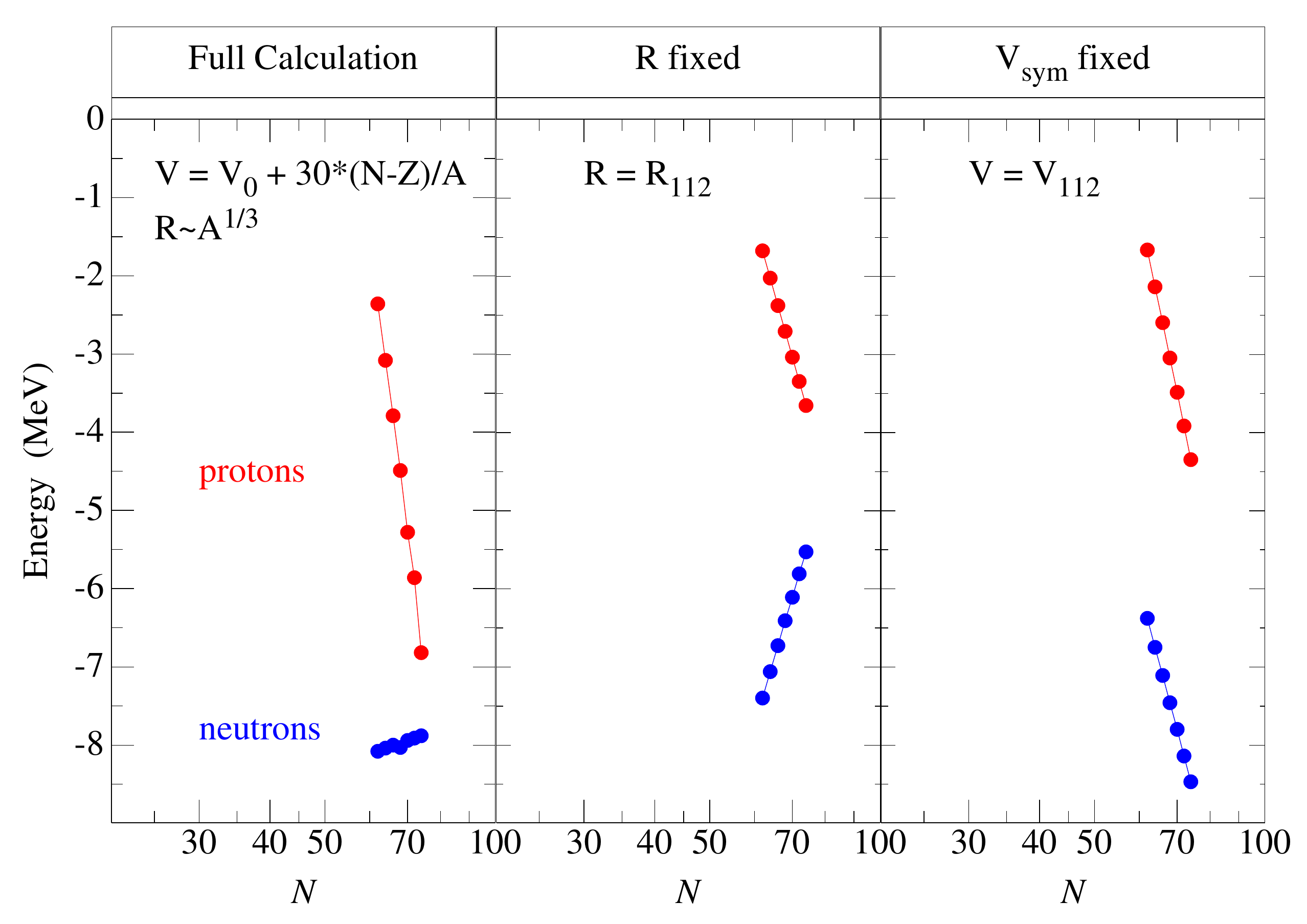}
\caption{\label{fig3} Calculations of the changing energy of a $1/2^+$ state in a Woods-Saxon potential for the seven stable even Sn isotopes. The red points show the behavior of a proton state, the blue ones for a neutron state. The left panel shows the behavior with neutron number $N$ with a typical potential, radius increasing as $A^{1/3}$ and a symmetry term proportional to $(N-Z)/A$.  The middle panel shows the results with the same symmetry term and the radius fixed at that for $^{112}$Sn.   The right hand panel is with no $(N-Z)/A$ dependence, the potential fixed at the value for $^{112}$Sn, and the radius again
changing as $A^{1/3}$.}
\end{figure}

As shown in Fig.~\ref{fig3}, a pattern qualitatively similar to that seen in the data requires a cancellation between the effect of the symmetry term and the changing radius.  A symmetry term with a fixed radius would yield a dependence with slopes that are roughly equal and opposite, while a changing radius with no symmetry term would give essentially the same slopes. But when both are included, the pattern is similar to what is seen in the data.

The near-cancellation of the effects of an increasing radius and a decreasing potential strength happens in the potential for neutrons with changing neutron number. But in the proton potential the radius still increases, but the sign of the symmetry-term contribution to the potential changes, the two terms have the same sign, and reinforce the change. That is as should be expected with an empirical potential.

One should not regard this as an `explanation,' rather that the empirical parameterization does describe the data, reasonably well. The symmetry term implicitly assumes that it is a perturbation on the main part of the interaction. But in view of this pattern that has been well known, and that we again demonstrate in the data collected in this paper, one could also write a mathematically exactly equivalent form in terms of a first term describing the interaction with the other type of nucleons $V_{np}$, and the second with the same type, $V_{T=1}$ instead of  $V_0$ and $V_{sym}$.  This would yield $V_{np}\cong-80$ MeV and $V_{T=1}\cong-20$ MeV (and thus, $V_{T=0}\cong-140$).  

The observation that the $T=0$ and `$np$' interactions are much stronger than that between identical $T=1$ nucleon pairs is certainly not new. The interaction between the increasing radius and the symmetry term, is built into the empirical parameterization of the Woods-Saxon potential.   The symmetry term also must account for the changing radius in a different way, and the cancellation between the two effects arises from the model used.  Consider the string of Sn isotopes with 50 protons.  These protons are the dominant source of the potential that binds the neutrons.  The closed shell of protons remains essentially the same throughout the Sn isotopes, except that the nuclei must be getting larger with added neutrons and the protons spread out over a larger volume.  If these protons contribute equally to the overall potential, then their dilution will correspond to a weaker potential.  This qualitative argument needs to be taken with caution, since the measured charge radius is changing in a somewhat complex way; at a rate slower than $A^{1/3}$ when neutrons are added, but the rate is not the same in the Sn isotopes as in the Ca isotopes. The use of the symmetry term in average potential is a useful empirical tool, but it is an approximation.

To discuss the two groups of slopes for the $T=1$ segments we consider the behavior of the corresponding two-body matrix elements from Ref.~\cite{Schiffer76} that are shown in the inset of Fig.~\ref{fig2}.  Here the behavior of the two groups of $T=1$ slopes is very similar to the pattern of $T=1$ matrix elements in ~\cite{Schiffer76} for the $j_1=j_2$ and the $j_1\neq j_2$ matrix elements. In that reference, the very small (and sometimes repulsive) matrix elements were approximately reproduced by two interactions, an attractive Yukawa potential with a range of one-pion exchange, and a longer range repulsive term, that tended to cancel the average attraction; the fit was not sensitive to the exact range of the repulsive term as long as it was longer than that of the attractive one.  The need and magnitude of the attraction was determined by the higher multipole content of the two-body spectra, in particular see the behavior of the $T=1$ matrix elements in Figs. 2 and 6 of Ref.~\cite{Schiffer76}.

In re-examining the details of that cancellation, it seems that the attractive term is slightly larger (by a few percent) when the two nucleons are in the same orbit, than when they are in different orbits. The data requires a near-zero monopole term. In Ref.~\cite{Schiffer76}, oscillator wave functions were used without allowing for a changing radius between, for instance, $^{41}$Ca and $^{42}$Ca. Using crude arguments, the magnitude of this effect of a changing radius is comparable to making the monopole interaction slightly repulsive.

In retrospect, and in view of the behavior with radius shown in Fig.~\ref{fig3}, one may wonder whether the apparent repulsive term may have its origin in the effect of the inevitably changing radius, that is traditionally ignored.  But our purpose here is to point out trends in the data, and leave it to further work to come up with possible model explanations for the simple trends that are apparent in the largely model-independent features.

To summarize, our survey of the data indicates at least four features that, at least, the present authors were not aware of, and that may help as a guide to our intuition, beyond what seems to be generally understood and accepted;

\begin{enumerate}

\item
The change in single-particle energies with nucleon number is qualitatively different, depending on whether the nucleons in the state are the same or different from those changing;

\item
The change per nucleon, is essentially the same as the experimentally observed effective $NN$ interaction;

\item
The change with {\it fractional} change in nucleon number is remarkably constant from $A\sim16$ to 208;

\item
For identical nucleons, the change is small, but it does matter whether the nucleons are all in the same orbit or not. The change in radius between adjacent nuclei, seems to have a significant influence on the effective interaction, particularly between identical nucleons.

\end{enumerate}

As the field moves toward more and more exotic nuclei, near the limits of binding, and the quality of data improves, the question of how the binding energies of individual orbits change with added nucleons, and exploring and understanding systematics is likely to become increasingly important.

We gratefully acknowledge stimulating conversations with many colleagues.  This material is based upon work supported by the U.S.\ Department of Energy, Office of Science, Office of Nuclear Physics, under Contract Number DE-AC02-06CH11357.


\end{document}